\documentclass{ws-ijmpa}

\begin{document}

\markboth{Rolando Gaitan} {On unitarity of a Yang-Mills
formulation for... }

%
\catchline{}{}{}{}{}
%

\title{ON UNITARITY OF A LINEARIZED YANG-MILLS FORMULATION FOR
MASSLESS AND MASSIVE GRAVITY WITH PROPAGATING TORSION}

\author{Rolando Gaitan Deveras}

\address{Departamento de F\'\i sica, Facultad de Ciencias y Tecnolog\'\i a, Universidad de
Carabobo,\\ A.P. 129 Valencia 2001, Edo. Carabobo,
Venezuela.\\
rgaitan@uc.edu.ve}

\maketitle

\begin{history}
\received{Day Month Year}
\revised{Day Month Year}
\end{history}

\begin{abstract}
A perturbative regime based on contortion as a dynamical variable
and metric as a (classical) fixed background, is performed in the
context of a pure Yang-Mills formulation for gravity in a $2+1$
dimensional space-time. In the massless case we show that the
theory contains three degrees of freedom and only one is a
non-unitary mode. Next, we introduce quadratical terms dependent
on torsion, which preserve parity and general covariance. The
linearized version reproduces an analogue
Hilbert-Einstein-Fierz-Pauli unitary massive theory plus three
massless modes, two of them represents non-unitary ones. Finally
we confirm the existence of a family of unitary
Yang-Mills-extended theories which are classically consistent with
Einstein's solutions coming from non massive and topologically
massive gravity. The unitarity of these YM-extended theories is
shown in a perturbative regime. A possible way to perform a
non-perturbative study is remarked.

\keywords{Gauge; Gravity; Unitarity.}
\end{abstract}

\ccode{PACS numbers: 04.50.Kd, 04.20.Fy, 04.60.Rt}

\section{Introduction}

There were some contributions on the exploration of classical
consistency of a pure Yang-Mills (YM) type formulation for
gravity, including the cosmological extension\cite{g1,g2} (and the
references therein), among others. In those references, Einstein's
theory is recovered after the imposition of torsion constraints.

Unfortunately, the path to a quantum version (if it is finally
possible) is not straightforward. For example, it is well known
that the Lagrangian of a pure YM theory based on the Lorentz group
$SO(3,1)\simeq SL(2,C)$\cite{kp} leads to a non-positive
Hamiltonian (due to non-compactness of the aforementioned gauge
group) and, then the canonical quantization procedure fails.
However, there is a possible way out if it is considered an
extension of the YM model thinking about a theory like
Gauss-Bonnet with torsion\cite{kpp} and this is confirmed because
the existence of a possible family of quadratical curvature
theories from which can be recovered unitarity\cite{sez}.

A first aim of this work is to expose, with some detail, a similar
(and obvious) situation about non-unitarity in a YM formulation
for gravity in both massless and massive theories. There is an
interest focussed in the study of massive gravity and propagating
torsion\cite{hvn}, among others. Particularly, the massive
versions that we will explore here arise, on one hand from some
quadratical terms set ($T^2$-terms) preserving parity which
depends on torsion (the old idea about considering $T^2$-terms in
a dynamical theory of torsion has been considered in the
past\cite{hs}) and, at a perturbative regime they give rise to a
Fierz-Pauli's massive term (in analogy with the recently BHT
model\cite{bht}). On the other hand, we remark some aspects of the
topologically masive version of the YM gravity\cite{g2} which do
not preserves parity and how is the way to reach unitarity in a
perturbative level, at least.

Whatever the model considered (massive or non massive), throughout
this work we follow the spirit of Kibble's idea\cite{kibble}
treating the metric as a fixed background, meanwhile the torsion
(contortion) will be considered as a dynamical field and it would
be thought as a quantum fluctuation around a classical fixed
background.

This paper is organized as follows. The next section is devoted to
a brief review on notation of the cosmologically extended YM
formulation\cite{g1} in $N$-dimensions and its topologically
massive version in $2+1$ dimension\cite{g2}. In section 3, we
consider the scheme of linearization of the massless theory around
a fixed Minkowskian background, allowing fluctuations on torsion.
Next, the Lagrangian analysis of constraints and construction of
the reduced action is performed, showing that this theory does
propagate degrees of freedom, including a ghost. We end this
section with some remarks about the counting of degrees of freedom
in a non perturbative level. In section 4, we introduce an
appropiate $T^2$-terms, which preserve parity, general covariance,
and its linearization gives rise to a Fierz-Pauli mass term.
There, the non-positive definite Hamiltonian problem gets worse:
the Lagrangian analysis shows that the theory has more non-unitary
degrees of freedom and we can't expect other thing. Gauge
transformations are explored in section 5. Although $T^2$-terms
provide mass only to some spin component of contortion, the
linearized theory loses the gauge invariance and there is no
residual invariance. This is clearly established through a
standard procedure for the study of possible chains of gauge
generators\cite{c}. In section 6 we confirm the well known fact
that there exists a family of theories which can cure the ghost
problem\cite{sez}, at least at perturbative level. Those theories
are classically consistent when it is shown that the set of
solutions contains the Einsteinian's ones. We end up with some
concluding remarks.

\section{A pure Yang-Mills formulation for gravity:
massless and topological massive cases}

Let $M$ be an $N$-dimensional manifold with a metric, $g_{\mu \nu
}$ and coordinates transformations, $U$ provided. A (principal)
fiber bundle is constructed with $M$ and a 1-form connection is
given, ${(A_\lambda)^\mu}_\nu $ which will be non metric
dependent. The affine connection transforms as ${A_\lambda}^\prime
=UA_\lambda U^{-1} + U
\partial _\lambda U^{-1}$ with $U \in GL(N,R)$. Torsion and curvature tensors are
${T^\mu}_{\lambda\nu}={(A_\lambda)^\mu}_\nu-{(A_\nu)^\mu}_\lambda$
and $F_{\mu\nu} \equiv \partial_\mu A_\nu - \partial_\nu A_\mu +
[A_\mu , A_\nu ]$, respectively. The contortion tensor is defined
by ${K^\lambda}_{\mu\nu}\equiv \frac
1{2}({T^\lambda}_{\mu\nu}+{{T_\mu}^\lambda}_\nu +
{{T_\nu}^\lambda}_\mu)$. Components of the Riemann-Cartan tensor
are ${R^\sigma}_{\alpha \mu \nu }\equiv ({F_{\nu \mu})^\sigma
}_\alpha$. The gauge invariant action with cosmological
contribution is\cite{g1}
\begin{equation}
{S^{(N)}}_0=\kappa^{2(4-N)} \big<-\frac 14 \, tr\, F^{\alpha
\beta}F_{\alpha \beta}+ q(N) \lambda^2 \big> \,\, , \label{eqa1}
\end{equation}
where $\kappa^2$ is in length units, $\big<...\big>\equiv \int d^N
x \sqrt{-g}(...) $, $\lambda$ is the cosmologic constant and the
parameter  $q(N)=2(4-N)/(N-2)^2(N-1)$ depends on dimension. The
shape of  $q(N)$ allows the recovering of (free) Einstein's
equations as a particular solution when the torsionless Lagrangian
constraints are imposed and $q(N)$ changes it sign when $N>5$. The
field equations are ${T_g}^{\alpha\beta}=
-\kappa^2g^{\alpha\beta}\lambda^2$ where
${T_g}^{\alpha\beta}\equiv
\kappa^2\,tr[F^{\alpha\sigma}{F^\beta}_\sigma
-\frac{g^{\alpha\beta}}4 \, F^{\mu \nu}F_{\mu \nu}]$ is the
energy-momentum tensor of gravity, and equation coming from
variation of connection is $\frac 1{\sqrt{-g}}\,\,\partial _\alpha
(\sqrt{-g}\,\,F^{\alpha \lambda}) + [A_\alpha , F^{\alpha
\lambda}] =0$, which can be rewritten as follows
\begin{eqnarray}
\nabla_\mu R_{\sigma\lambda}-\nabla_\lambda R_{\sigma\mu} =0\,\,
,\,\, \label{g1a}
\end{eqnarray}
and the trace $\sigma-\lambda$ gives the expected condition
$R=constant$.

It is well known that the introduction of a Chern-Simons
lagrangian term (CS) in the Hilbert-Einstein formulation of
gravity provides a theory which describes a massive excitation of
a graviton in 2+1 dimensions\cite{DJT}. If a cosmological term is
included, the cosmologically extended topological massive gravity
(TMG$\lambda$) arises\cite{d}.

So, the study of consistence of a Yang-Mills type formulation for
topological massive gravity with cosmological constant has been
performed\cite{g2}. There, it is verifying the existence of causal
propagation and the fact that the standard TMG$\lambda$ can be
recovered from the aforementioned model at the torsionless limit.
The model is
\begin{equation}
S={S^{(3)}}_0 +
\frac{m\kappa^2}{2}\big<\varepsilon^{\mu\nu\lambda}\,tr\big( A_\mu
\partial_\nu A_\lambda +\frac{2}{3}\, A_\mu A_\nu A_\lambda \big)\big>\,
\, , \label{eqa2}
\end{equation}
which does not preserve parity and ${S^{(3)}}_0$ is given by
(\ref{eqa1}) for $N=3$.

Now, the torsionless limit of (\ref{eqa2}) is explored by
introducing nine torsion's constraints through the new action $S'
= S+\kappa^2\int d^3 x \sqrt{-g}
\,b_{\alpha\beta}\,\varepsilon^{\beta\lambda\sigma}{(A_\lambda)^\alpha}_\sigma$,
where the nine auxiliary fields $b_{\alpha\beta}$ can be seen as
Lagrange multipliers. Variation on connection and metric gives
rise the following field equations
\begin{equation}
\nabla_\mu R_{\sigma\lambda}-\nabla_\lambda R_{\sigma\mu}
-m\,{\varepsilon^{\nu\rho}}_\sigma(g_{\lambda\nu}R_{\mu\rho}-g_{\mu\nu}R_{\lambda\rho}
-\frac{2}{3}\,Rg_{\lambda\nu}g_{\mu\rho})=0\,\, , \label{eq13}
\end{equation}
\begin{equation}
R_{\sigma\mu}{R^\sigma}_\nu
-RR_{\mu\nu}+\frac{g_{\mu\nu}}{4}\,R^2-g_{\mu\nu}\lambda^2=0\,\, ,
\label{eqa14}
\end{equation}
and Lagrange multipliers are
\begin{equation}
b_{\mu\nu}=\frac{mR}{6}\,g_{\mu\nu}\,\, . \label{eqa14a}
\end{equation}

The trace $\sigma-\lambda$ of (\ref{eq13}) leads to the following
consistency condition
\begin{equation}
R=constant \,\, , \label{eqam10}
\end{equation}
and due to this condition on the Ricci scalar, we can test
particular solutions of the type
$R_{\mu\nu}=\frac{R}{3}\,g_{\mu\nu}$, by pluging them in
(\ref{eqa14}), and this gives
\begin{eqnarray}
R=\pm 6\mid\lambda\mid\,\, , \label{eqam13}
\end{eqnarray}
verifying the existence of (Anti) de Sitter solutions.

Finally, at the torsionless limit, the TMG$\lambda$ model is
recovered from this YM one if we take the mass value $m$ as the
mass of the Chern-Simons model and the consistency condition
(\ref{eqam10}) is fixed as (\ref{eqam13}).

\section{Linearization of the massless theory}

With a view on the performing of a perturbative study of the
massive model, we wish to note some aspects of the variational
analysis of free action (\ref{eqa1}) in $2+1$ dimensions. As we
had said above, the connection will be considered as  a dynamical
field whereas the space-time metric would be a fixed background,
in order to explore (in some sense) the isolated behavior of
torsion (contortion) and avoid higher order terms in the field
equations. For simplicity we assume $\lambda =0$.

Then, let us  consider a Minkowskian space-time with a metric
$diag(-1,1,1)$ provided and, obviously with no curvature nor
torsion. The notation is
\begin{equation}
\overline{g}_{\alpha\beta}=\eta_{\alpha\beta}\, \, , \label{eqb1}
\end{equation}
\begin{equation}
{\overline{F}}^{\alpha\beta}=0\, \, , \label{eqb2}
\end{equation}
\begin{equation}
{\overline{T}^\lambda}_{\mu\nu}=0\, \, . \label{eqb3}
\end{equation}
It can be observed that curvature ${\overline{F}}^{\alpha\beta}=0$
and torsion ${\overline{T}^\lambda}_{\mu\nu}=0$, in a space-time
with metric $\overline{g}_{\alpha\beta}=\eta_{\alpha\beta}$
satisfy the background equations, $\frac
1{\sqrt{-\overline{g}}}\,\,\partial _\alpha
(\sqrt{-\overline{g}}\,\,\overline{F}^{\alpha \lambda}) +
[\overline{A}_\alpha , \overline{F}^{\alpha \lambda}] =0$ and
${\overline{T}_g}^{\alpha\beta}=0$, identically.

Thinking in variations
\begin{equation}
A_\mu=\overline{A}_\mu+ a_\mu\,\,\,\,,\,\,\,\,|a_\mu|\ll 1\, \, ,
\label{eqb4}
\end{equation}
for this case $\overline{A}_\mu=0$. Then,  action (\ref{eqa1})
takes the form
\begin{equation}
{S^{(3)L}}_0 =\kappa^2 \big<-\frac 14 \, tr\, f^{\alpha
\beta}(a)f_{\alpha \beta}(a)\big> \,\, , \label{eqb5}
\end{equation}
where $f_{\alpha \beta}(a)=\partial_\alpha a_\beta-\partial_\beta
a_\alpha $ and (\ref{eqb5}) is gauge invariant under
\begin{equation}
\delta a_\mu = \partial_\mu \omega  \,\, , \label{eqb6}
\end{equation}
with $\omega \in G=SO(1,2)$.

In order to describe in detail the action (\ref{eqb5}), let us
consider the following decomposition for perturbed connection
\begin{equation}
{(a_\mu)^{\alpha}}_\beta={\epsilon^{\sigma\alpha}}_\beta
k_{\mu\sigma}+{\delta^\alpha}_\mu v_\beta
-\eta_{\mu\beta}v^\alpha\,\, , \label{eqb7}
\end{equation}
where $k_{\mu\nu}=k_{\nu\mu}$ and $v_\mu$ are the symmetric and
antisymmetric parts of the rank two perturbed contortion  (i. e.,
the rank two contortion is $K_{\mu\nu}\equiv -\frac{1}2
{\epsilon^{\sigma\rho}}_\nu K_{\sigma\mu\rho}$), respectively. It
can be noted that decomposition  (\ref{eqb7}) has not been
performed in irreducible spin components and explicit writing down
of the traceless part of $k_{\mu\nu}$ would be needed. This
component will be considered when the study of reduced action will
be performed. Using (\ref{eqb7}) in (\ref{eqb5}), we get
\begin{equation}
{S^{(3)L}}_0 =\kappa^2 \big< k_{\mu\nu}\Box{}k^{\mu\nu}+
\partial_\mu k^{\mu\sigma}\partial_\nu {k^\nu}_\sigma
-2 \epsilon^{\sigma\alpha\beta}\partial_\alpha v_\beta\partial_\nu
{k^\nu}_\sigma - v_\mu \Box{}v^\mu +(\partial_\mu v^\mu)^2\big>
\,\, , \label{eqb8}
\end{equation}
which is gauge invariant under the following transformation rules
(induced by (\ref{eqb6}))
\begin{equation}
\delta k_{\mu\nu}= \partial_\mu \xi_\nu + \partial_\nu \xi_\mu
\,\, , \label{eqb9}
\end{equation}
\begin{equation}
\delta v_\mu =-{\epsilon^{\sigma\rho}}_\mu \partial_\sigma
\xi_\rho \,\, , \label{eqb10}
\end{equation}
with $\xi_\mu \equiv \frac{1}4
{\epsilon^\beta}_{\alpha\mu}{w^\alpha}_\beta$. These
transformation rules clearly show that only the antisymmetric part
of $w$ is needed (i. e.: only three gauge fixation would be
chosen).

Now, let us study the system of Lagrangian constraints in order to
explore the number of degrees of freedom. A possible approach
consists in a $2+1$ decomposition of the action (\ref{eqb8}) in
the way
\begin{eqnarray}
{S^{(3)L}}_{0} =\kappa^2
\big<[-\dot{k}_{0i}+2\partial_ik_{00}-2\partial_nk_{ni}-2\epsilon_{in}\dot{v}_n
+2\epsilon_{in}\partial_n v_0 ]\dot{k}_{0i}\nonumber
\\+\dot{k}_{ij}\dot{k}_{ij}
+[2\epsilon_{nj}\partial_nk_{00}+2\epsilon_{nj}\partial_mk_{nm}-\dot{v}_j-2\partial_jv_0]\dot{v}_j\nonumber
\\+2(\dot{v}_0)^2 +k_{00}\Delta k_{00}-2k_{0i}\Delta
k_{0i}+k_{ij}\Delta k_{ij}-(\partial_ik_{i0})^2\nonumber
\\+\partial_nk_{ni}\partial_mk_{mi} -2\epsilon_{ij}\partial_i v_j
\partial_nk_{n0}-2\epsilon_{lm}\partial_m v_0
\partial_nk_{nl}+\nonumber \\v_0\Delta v_0-v_i\Delta v_i +(\partial_nv_n)^2\big>
\label{eqb14a}
\end{eqnarray}
and using a Transverse-Longitudinal (TL) decomposition\cite{pio}
with notation
\begin{eqnarray}
k_{00}\equiv n \, \, , \label{a1}
\end{eqnarray}
\begin{eqnarray}
h_{i0}=h_{0i} \equiv \partial_i k^L + \epsilon_{il}\partial_l
k^T\, \, , \label{a2}
\end{eqnarray}
\begin{eqnarray}
k_{ij}=k_{ji}\equiv (\eta_{ij}\Delta -\partial _i \partial
_j)k^{TT} +\partial _i \partial _j k^{LL}\nonumber
\\ +(\epsilon _{ik}\partial
_k\partial _j +\epsilon _{jk}\partial _k \partial _i )k^{TL}
\label{a3}
\end{eqnarray}
\begin{eqnarray}
v_0\equiv q \, \, , \label{a4}
\end{eqnarray}
\begin{eqnarray}
v_i \equiv \partial_i v^L + \epsilon_{il}\partial_l v^T\, \, ,
\label{a5}
\end{eqnarray}
where $\Delta \equiv \partial _i \partial _i$, eq. (\ref{eqb14a})
can be rewritten as follows
\begin{eqnarray}
{S^{(3)L}}_{0} =\kappa^2 \big<\dot{k}^L \Delta \dot{k}^L+\dot{k}^T
\Delta \dot{k}^T+\dot{v}^L \Delta \dot{v}^L+ \dot{v}^T \Delta
\dot{v}^T\nonumber \\+2\dot{v}^L \Delta \dot{k}^T-2\dot{v}^T
\Delta \dot{k}^L +(\Delta \dot{k}^{TT})^2 +(\Delta \dot{k}^{LL})^2
\nonumber \\+2(\Delta \dot{k}^{TL})^2+2(\dot{q})^2-2n\Delta
\dot{k}^L +2n\Delta \dot{v}^T\nonumber \\+2q\Delta
\dot{v}^L-2q\Delta \dot{k}^T+2\Delta k^{LL}\Delta
\dot{k}^L+2\Delta k^{TL}\Delta \dot{k}^T\nonumber \\+2\Delta
k^{LL}\Delta \dot{v}^T -2\Delta k^{TL}\Delta \dot{v}^L +q\Delta
q+n\Delta n\nonumber \\+(\Delta k^L)^2+2(\Delta k^T)^2 +2(\Delta
v^L)^2 +(\Delta v^T)^2\nonumber \\+2\Delta v^{T}\Delta k^{L}
+2q\Delta^2 k^{TL} +\Delta k^{TT}\Delta^2 k^{TT}\nonumber
\\+\Delta k^{TL}\Delta^2 k^{TL}\big>
\,\,. \label{e14}
\end{eqnarray}

Primary Lagrangian constraints, joined to some links among
accelerations, can be obtained through an inspection on field
equations, which arise from (\ref{e14}). A ''Coulomb'' gauge
(i.e., $\partial_ik_{i\mu}=0$) is considered. Then, we get the
following set of twelve Lagrangian constraints
\begin{equation}
n=\dot{n}=v^T=\dot{v}^T=k^L=\dot{k}^L=k^{LL}=\dot{k}^{LL}=k^{TL}=\dot{k}^{TL}=0
\,\, , \label{e32}
\end{equation}
\begin{equation}
\dot{k}^T-\dot{v}^L+q=0 \,\, , \label{e33}
\end{equation}
\begin{equation}
\Delta k^T-\Delta v^L+\dot{q}=0 \,\, , \label{e34}
\end{equation}
and all accelerations are solved. Then, there are three degrees of
freedom, and the constraint system give rise to reduced action
\begin{eqnarray}
{S^{(3)L*}}_{0} =\kappa^2 \big<4\dot{k}^T \Delta
\dot{k}^T+4(\Delta k^T)^2 +4(\dot{q})^2\nonumber \\+4q\Delta q
+(\Delta \dot{k}^{TT})^2+\Delta k^{TT}\Delta^2 k^{TT}\big> \,\,.
\label{e35}
\end{eqnarray}
Introducing notation
\begin{equation}
Q\equiv 2q \,\, , \label{e36}
\end{equation}
\begin{equation}
Q^T\equiv 2(-\Delta)^{\frac{1}2}k^T \,\, , \label{e37}
\end{equation}
\begin{equation}
Q^{TT}\equiv \Delta k^{TT} \,\, , \label{e38}
\end{equation}
the reduced action is rewritten as follows
\begin{eqnarray}
{S^{(3)L*}}_{0} =\kappa^2 \big<Q\Box{}Q-Q^T\Box{}Q^T+
Q^{TT}\Box{}Q^{TT}\big>\,\,, \label{e39}
\end{eqnarray}
showing two unitary and one non-unitary modes, then the
Hamiltonian is not positive definite. This study could also have
considered from the point of view of the exchange amplitude
procedure, in which is considered the coupling to a (conserved)
energy-momentum tensor of some source, trough Lagrangian terms
$\kappa k_{\mu\nu}T^{\mu\nu}$ and $\chi v_{\mu}J^{\mu}$.

Some features about the degrees of freedom's counting of this
model at a non perturbative regime can be remarked. If one keep in
mind a physical system where the metric is considered as a non
dynamical field (otherwise we get a new problem with additional
degrees of freedom coming from fluctuations of the metric), the
essencial problem to face up here is related to the Lagrangian (or
Hamiltonian) analysis of a second order action with potentials
which depend on the fourth-order power of the contortion field. In
detail, the Riemann-Cartan curvature, $({F_{\mu \nu}(A))^\sigma
}_\alpha $ can be decomposed in terms of the Riemann-Christofell,
$({F_{\mu \nu}(\Gamma))^\sigma }_\alpha $ and contortion, this
means
\begin{eqnarray}
F_{\mu \nu}(A)=F_{\mu \nu}(\Gamma)+F_{\mu \nu}(K)+[\Gamma_\mu,
K_\nu]+[K_\mu, \Gamma_\nu]\,\,, \label{e40}
\end{eqnarray}
where the components of the matrix $\Gamma_\mu$ and $K_\mu$ are
the Christofell's symbols and contortion components, respectively.
Then, the action (\ref{eqa1}) for $N=3$ and $\lambda=0$ is
rewritten for any given metric as follows
\begin{eqnarray}
{S^{(3)}}_0=\kappa^{2} tr\,\big<-\frac 14 \,  F^{\alpha
\beta}(K)F_{\alpha \beta}(K) +{P_1}^{\alpha
\beta}(K,\Gamma)F_{\alpha \beta}(K) +{P_2}(K,\Gamma)\big> \,\, ,\nonumber \\
\label{e41}
\end{eqnarray}
where ${P_1}^{\alpha \beta}(K,\Gamma)$ and ${P_2}(K,\Gamma)$ are
polynomials of order one and two in contortion, respectively and
they are identically null when $\Gamma_\mu =0$. One can start
considering a space-time with a Minkowskian metric provided (i.
e., a Weitzenb$\ddot{o}$ck space), hence action (\ref{e41}) is
rewritten as
\begin{eqnarray}
{S^{(3)}}_{W0}=\kappa^{2} tr\,\big<-\frac 14 F^{\alpha
\beta}(K)F_{\alpha \beta}(K)\big> \,\, , \label{e42}
\end{eqnarray}
which perturbative regime has been studied above. Next, we will
show that this theory contains three degrees of freedom as in the
linearized level. For this purpose, there is a way inspired in the
well known first order formalism for Yang-Mills
theories\cite{mckeon}, reducing derivatives and potential's
powers, simultaneously. In $2+1$ dimension, particularly one can
consider a rank three auxiliary field with notation
${(f_\mu)^\alpha}_\beta$. Then, the first order version of the
action (\ref{e42}) is introduced through $S_{W0}=\kappa^{2}
tr\,\big<\frac 12 \, f^\mu f_\mu- \frac{\epsilon ^{\mu\nu\lambda}
}{2}\,f_\lambda F_{\mu\nu}(K) \big>$. After its $2+1$ splitting
and using Lagrangian constraints (removing $K_0$ and $f_0$), the
reduced action is
\begin{equation}
S^*_{W0}=\kappa^{2} tr\,\big<\epsilon_{ij} \,f_i
\dot{K}_j+\frac{f_if_i}{2} \big> \,\, , \label{e44}
\end{equation}
where $\epsilon_{ij}\equiv {\epsilon^0}_{ij}$. The Lagrangian
constraints joined with the ''Coulomb'' gauge fixation (i. e.,
$\partial_i K_i=0$) constitute a set of forty two constraints
\begin{equation}
\epsilon_{ij} \,\dot{K}_j+f_i=0 \,\, , \label{e44a}
\end{equation}
\begin{equation}
\dot{f}_j=0 \,\, , \label{e44a}
\end{equation}
\begin{equation}
\partial_i K_i=0 \,\, , \label{e44a}
\end{equation}
\begin{equation}
\partial_i \dot{K}_i=0\,\, , \label{e44a}
\end{equation}
for forty eight fields and velocities, confirming the existence of
three degrees of freedom.

The main question is about the behavior of the theory for any
given metric. Again we resort to auxiliary fields $f_\mu$ and the
first order version of the action (\ref{eqa1}) for $N=3$ and
$\lambda = 0$, is $S_0=\kappa^{2} tr\,\big<\frac 12 \, f^\mu
f_\mu- \frac{\varepsilon ^{\mu\nu\lambda} }{2}\,f_\lambda
F_{\mu\nu}(A) \big>$. Using (\ref{e40}), this action can be
written in explicit terms of the dynamical variables
\begin{equation}
S_0=\kappa^{2} tr\,\big<\frac 12 \, f^\mu f_\mu - Q^\mu(g) f_\mu-
\varepsilon ^{\mu\nu\lambda} \,f_\lambda \big( \frac12
F_{\mu\nu}(K)+[\Gamma_\mu , K_\nu]\big) \big> \,\, , \label{e45}
\end{equation}
where $Q^\sigma(g)$ is Poincare's dual of Riemann-Christofell's
curvature given by
\begin{equation}
Q^\sigma(g)\equiv \frac{\varepsilon ^{\mu\nu\sigma}
}{2}\,F_{\mu\nu}(\Gamma)  \,\, , \label{e46}
\end{equation}
this means, ${\big(Q^\sigma(g) \big)^\alpha}_\beta =\varepsilon
^{\mu\nu\sigma}\big( {\delta^\alpha}_\nu
R_{\beta\mu}-g_{\beta\nu}{R^\alpha}_\mu-
\frac{R}2\,{\delta^\alpha}_\nu g_{\beta\mu} \big) $.

It is possible to show, under certain conditions a narrow analogy
between this theory and that of the Weitzenb$\ddot{o}$ck, given by
action (\ref{e44}). Thinking about the antisymmetric property of
the contortion ($K^{\alpha\mu\nu}=-K^{\nu\mu\alpha}$), the first
step is to consider a symmetric-antisymmetric decomposition of all
fields involved in the action (\ref{e45}). In other words, let us
introduce the next decomposition
\begin{equation}
\Gamma_\mu=\overline{\Gamma}_\mu+\widetilde{\Gamma}_\mu \,\, ,
\label{e47}
\end{equation}
\begin{equation}
f_\mu=\overline{f}_\mu+\widetilde{f}_\mu \,\, , \label{e48}
\end{equation}
where $\overline{\Gamma}_\mu$ and $\widetilde{\Gamma}_\mu$ are the
symmetric and antisymmetric parts of the Christofell's symbols,
repectively (i. e., $(\overline{\Gamma}_\mu)_{\sigma\rho}=\frac12
\partial_\mu g_{\sigma\rho}$ and
$(\widetilde{\Gamma}_\mu)_{\sigma\rho}=\frac12(
\partial_\rho g_{\sigma\mu}-\partial_\sigma g_{\rho\mu})$). The
same idea is reflected in notation of (\ref{e48}). Using these
definitions in (\ref{e45}) and performing a $2+1$ splitting, the
set of Lagrange constraints allow us to remove the non dynamical
fields (this means $\widetilde{f}_0$ and $\overline{f}_\mu$), then
the reduced action is
\begin{equation}
S^*_0=\kappa^{2} tr\,\big<\varepsilon^{ij} \,\widetilde{f}_i
\big(\dot{K}_j+[\widetilde{\Gamma}_0 , K_j]-\mathcal{Q}_j(g) \big)
+\frac{g^{ij}}{2} \widetilde{f}_i\widetilde{f}_j\big> \,\, .
\label{e49}
\end{equation}
with $Q^i(g)=\varepsilon^{ij}\mathcal{Q}_i(g)$.

Next, a new antisymmetric variable (in the sense of definitions
(\ref{e47}) and (\ref{e48})) is introduced
\begin{equation}
q_j= K_j-{\mathcal{Q}^*}_j(g) \,\, , \label{e49a}
\end{equation}
where ${\mathcal{Q}^*}_j(g)$ is a solution of the non homogeneous
and first order differential equation
$\dot{\mathcal{Q}^*}_i(g)+[\widetilde{\Gamma}_0 ,
{\mathcal{Q}^*}_i(g)]=\mathcal{Q}_i(g)$. This suggests the
definition of the following operator which acts on any object with
matrix representation (i. e., a subgroup of $GL(3,R)$), $\chi$, in
other words
\begin{equation}
\widetilde{\nabla}_\mu \chi\equiv \partial_\mu \chi
+[\widetilde{\Gamma}_\mu , \chi] \,\, , \label{e50}
\end{equation}
and on a real function $h$
\begin{equation}
\widetilde{\nabla}_\mu h \equiv \partial_\mu h \,\, , \label{e50a}
\end{equation}
for example $\widetilde{\nabla}_\mu \varepsilon^{ij}=\partial_\mu
\varepsilon^{ij}$, $\widetilde{\nabla}_\mu g^{ij}=\partial_\mu
g^{ij}$, etc.

Let $\chi$ and $\xi$ be objects with matrix representation, some
properties of $\widetilde{\nabla}_\mu$ are
\begin{equation}
\widetilde{\nabla}_\mu (\xi\chi)= \xi\widetilde{\nabla}_\mu
\chi+(\widetilde{\nabla}_\mu \xi)\chi \,\, , \label{e50b}
\end{equation}
\begin{equation}
[\widetilde{\nabla}_\mu,
\widetilde{\nabla}_\nu]\chi=[F_{\mu\nu}(\widetilde{\Gamma}) ,\chi]
\,\, . \label{e50c}
\end{equation}
It must be pointed out that $\widetilde{\nabla}_\mu$ is not a
covariant derivative for an arbitrary background.

Using (\ref{e49a}) and (\ref{e50}) in (\ref{e49}), we write the
reduced action as follows
\begin{equation}
S^*_0=\kappa^{2} tr\,\big<\varepsilon^{ij} \,\widetilde{f}_i
\widetilde{\nabla}_0 q_j+\frac{g^{ij}}{2}
\widetilde{f}_i\widetilde{f}_j\big> \,\, . \label{e51}
\end{equation}
and it gives rise a set of twelve constraints
\begin{equation}
\phi_i\equiv \widetilde{\nabla}_0 \widetilde{f}_i=0 \,\, ,
\label{e51a}
\end{equation}
\begin{equation}
\psi_i\equiv \widetilde{\nabla}_0 q_i
+{\varepsilon_{0i}}^j\widetilde{f}_j=0 \,\, , \label{e51b}
\end{equation}
whose preservation give the relations for accelerations
\begin{equation}
\dot{\phi}_i\simeq {\widetilde{\nabla}_0}^2 \widetilde{f}_i=0 \,\,
, \label{e52a}
\end{equation}
\begin{equation}
\dot{\psi}_i\simeq {\widetilde{\nabla}_0}^2 q_i
+\partial_0{\varepsilon_{0i}}^j\widetilde{f}_j=0 \,\,
.\label{e52b}
\end{equation}

However, the complete Lagrangian analysis depends on a gauge
fixation. In order to illustrate the remaining Lagrangian process
we find some similarities with the Weitzenb$\ddot{o}$ck case if
certain conditions are demanded on the background. Therefore, let
the metric be static-stationary, this means
$\partial_0g_{\mu\nu}=0$ and $g_{0i}=0$ (i. e., Schwarszchild
background), for instance relation (\ref{e52b}) gives
${\widetilde{\nabla}_0}^2 q_i =0 $.

Now, one can explore a gauge fixation. For example, the axial
gauge provide six additional constraints
\begin{equation}
\varphi^A \equiv q_2 =0 \,\, ,\label{e53a}
\end{equation}
\begin{equation}
\dot{\varphi}^A\simeq\widetilde{\nabla}_0 q_2 =0 \,\,
,\label{e53b}
\end{equation}
and joined to (\ref{e51a}) and (\ref{e51b}), say that there are
three degrees of freedom. An equivalent procedure can be developed
if one perform a ''Coulomb'' gauge fixation
\begin{equation}
\varphi^C \equiv \widetilde{\nabla}_i q_i + [a_i , q_i]=0 \,\,
,\label{e54a}
\end{equation}
where $a_i$ satisfies the differencial equation
$\widetilde{\nabla}_0 a_i=F_{i0}(\widetilde{\Gamma})$.
Preservation of (\ref{e54a}) provide another three constraints
\begin{equation}
\dot{\varphi}^C \simeq \widetilde{\nabla}_i
\widetilde{\nabla}_0q_i + [a_i , \widetilde{\nabla}_0q_i]=0 \,\,
,\label{e54b}
\end{equation}
and the procedure is finished (preservation of (\ref{e54b}) is
identically satisfied). Again, the constraint system shows three
degrees of freedom.

\section{YM gravity with parity preserving massive term}

It can be possible to write down a massive version which respect
parity and we introduce a possible model as follows
\begin{equation}
{S^{(3)}}_{m}={S^{(3)}}_0 - \frac{m^2
\kappa^2}{2}\big<{T^{\sigma}}_{\sigma
\nu}{{T^{\rho}}_{\rho}}^{\nu}-T^{\lambda\mu\nu}T_{\mu\lambda\nu}-\frac{1}{2}T^{\lambda\mu\nu}T_{\lambda\mu\nu}\big>\,
\, . \label{eqa3}
\end{equation}
In a general case, two types of field equations can be obtained if
independent variations on metric and connection are allowed. On
one hand, variations on metric give rise to the expression of the
gravitacional energy-momentum tensor, ${T_g}^{\alpha\beta}\equiv
\kappa^2\,tr[F^{\alpha\sigma}{F^\beta}_\sigma
-\frac{g^{\alpha\beta}}4 \, F^{\mu \nu}F_{\mu \nu}]$, in other
words
\begin{equation}
{T_g}^{\alpha\beta}=
-{T_t}^{\alpha\beta}-\kappa^2g^{\alpha\beta}\lambda^2\, \, ,
\label{eqa4}
\end{equation}
where ${T_t}^{\alpha\beta}\equiv
-m^2\kappa^2[3t^{\alpha\sigma}{t^\beta}_\sigma
+3t^{\sigma\alpha}{t_\sigma}^\beta
-t^{\alpha\sigma}{t_\sigma}^\beta-t^{\sigma\alpha}{t^\beta}_\sigma
- (t^{\alpha\beta}+t^{\beta\alpha}){t_\sigma}^\sigma
-\frac{5g^{\alpha\beta}}2 \, t^{\mu \nu}t_{\mu
\nu}+\frac{3g^{\alpha\beta}}2 \, t^{\mu \nu}t_{\nu
\mu}+\frac{g^{\alpha\beta}}2\,({t_\sigma}^\sigma)^2 ]$ is the
torsion contribution to the energy-momentum distribution and
$t^{\alpha\beta}\equiv\frac{\varepsilon^{\mu\nu\alpha}}2\,{T^{\beta}}_{\mu\nu}
$. This says, for example, that the quest of possible black hole
solutions must reveal a dependence on parameters  $m^2$ and
$\lambda^2$.

On the other hand, variations on connection provide the following
equations
\begin{eqnarray}
\frac 1{\sqrt{-g}}\,\,\partial _\alpha (\sqrt{-g}\,\,F^{\alpha
\lambda}) + [A_\alpha , F^{\alpha \lambda}] =J^\lambda \, \, ,
\label{eqa5}
\end{eqnarray}
where the current is $(J^\lambda)^{\nu}\,_\sigma =
m^2({\delta^\lambda}_\sigma {{K^\rho}_\rho}^\nu
-{\delta^\nu}_\sigma {{K^\rho}_\rho}^\lambda
+2{{K^\nu}_\sigma}^\lambda)$. We can observe in (\ref{eqa5}) that
contortion and metric appear as sources of gravity, where the
cosmological contribution is obviously hide in space-time metric.
In a weak torsion regime, equation (\ref{eqa5}) takes a familiar
shape, this means $\nabla_\alpha F^{\alpha \lambda} =J^\lambda $.

Now we explore the perturbation of the massive case given at
(\ref{eqa3}) and with the help of (\ref{eqb7}), the linearized
action is
\begin{eqnarray}
{S^{(3)L}}_m =\kappa^2 \big< k_{\mu\nu}\Box{}k^{\mu\nu}+
\partial_\mu k^{\mu\sigma}\partial_\nu {k^\nu}_\sigma
-2 \epsilon^{\sigma\alpha\beta}\partial_\alpha v_\beta
\partial_\nu {k^\nu}_\sigma\nonumber \\ - v_\mu \Box{}v^\mu +(\partial_\mu
v^\mu)^2-m^2(k_{\mu\nu}k^{\mu\nu}-k^2)\big> \,\, . \label{eqb15}
\end{eqnarray}

Using a TL-decomposition defined by (\ref{a1})-(\ref{a5}), we can
write  (\ref{eqb15}) in the way
\begin{eqnarray}
{S^{(3)L}}_m =\kappa^2 \big<\dot{k}^L \Delta \dot{k}^L+\dot{k}^T
\Delta \dot{k}^T+\dot{v}^L \Delta \dot{v}^L+ \dot{v}^T \Delta
\dot{v}^T+2\dot{v}^L \Delta \dot{k}^T-2\dot{v}^T \Delta \dot{k}^L
\nonumber \\
+(\Delta \dot{k}^{TT})^2 +(\Delta \dot{k}^{LL})^2 +2(\Delta
\dot{k}^{TL})^2+2(\dot{q})^2-2n\Delta \dot{k}^L +2n\Delta
\dot{v}^T\nonumber
\\+2q\Delta \dot{v}^L-2q\Delta
\dot{k}^T+2\Delta k^{LL}\Delta \dot{k}^L+2\Delta k^{TL}\Delta
\dot{k}^T+2\Delta k^{LL}\Delta \dot{v}^T \nonumber \\-2\Delta
k^{TL}\Delta \dot{v}^L +q\Delta q+n\Delta n+(\Delta
k^L)^2+2(\Delta k^T)^2 +2(\Delta v^L)^2\nonumber \\ +(\Delta
v^T)^2+2\Delta v^{T}\Delta k^{L} +2q\Delta^2 k^{TL} +\Delta
k^{TT}\Delta^2 k^{TT}+\Delta k^{TL}\Delta^2 k^{TL}\nonumber
\\+m^2[-2k^L\Delta k^L-2k^T\Delta k^T-2(\Delta  k^{TL})^2 -2n(\Delta  k^{TT}+\Delta  k^{LL})\nonumber
\\+2\Delta  k^{TT}\Delta  k^{LL}]\big>\,\,. \label{eqba1}
\end{eqnarray}
Here, there is no gauge freedom (as it will be confirmed in next
section) and field equations provide primary constraints and some
accelerations. The preservation procedure gives rise to a set of
eight constraints
\begin{equation}
\dot{v}^T-\dot{k}^L+n -m^2(k^{TT}+k^{LL})=0\,\, , \label{c1}
\end{equation}
\begin{equation}
\Delta \dot{k}^{LL}-\Delta k^L-\Delta v^T +m^2k^L=0\,\, ,
\label{c2}
\end{equation}
\begin{equation}
\Delta \dot{k}^{TL}-\dot{q}-\Delta k^T+\Delta v^L +m^2k^T=0 \,\, ,
\label{c3}
\end{equation}
\begin{equation}
\Delta \dot{k}^{LL}-\Delta k^L-\Delta v^T
+m^2(\dot{k}^{TT}+\dot{k}^{LL})=0\,\, , \label{c4}
\end{equation}
\begin{equation}
\dot{k}^L+\Delta k^{TT}-n =0 \,\, , \label{c5}
\end{equation}
\begin{equation}
\dot{k}^T-\Delta k^{TL}=0\,\, , \label{c6}
\end{equation}
\begin{equation}
\dot{v}^T+\Delta k^{TT}+m^2(k^{TT}+k^{LL})-2m^2\Delta^{-1} n=0
\,\, , \label{c7}
\end{equation}
\begin{equation}
\dot{n}-\Delta k^{L}=0 \,\, , \label{c8}
\end{equation}
which says that this massive theory get five degrees of freedom.
In order to explore the physical content, we can take a short path
to this purpose and it means to start with a typical
transverse-traceless (Tt) decomposition instead the
TL-decomposition one. Notation for the Tt-decomposition of fields
is
\begin{equation}
k_{\mu\nu}={k^{Tt}}_{\mu\nu}+\hat{\partial}_\mu {\theta^{T}}_\nu
+\hat{\partial}_\nu
{\theta^{T}}_\mu+\hat{\partial}_\mu\hat{\partial}_\nu \psi
+\eta_{\mu\nu}\phi \,\, , \label{eqb16}
\end{equation}
\begin{equation}
v_\mu={v^{T}}_\mu+\hat{\partial}_\mu v \,\, , \label{eqb17}
\end{equation}
with the subsidiary conditions
\begin{equation}
{k^{Tt\mu}}_\mu=0\,\,\,\,,\,\,\,\,\partial^\mu{k^{Tt}}_{\mu\nu}=0\,\,\,\,,\,\,\,\,\partial^\mu
{\theta^{T}}_\mu =0\,\,\,\,,\,\,\,\,\partial^\mu {v^{T}}_\mu =0
\,\, , \label{eqb18}
\end{equation}
where we use the operator $\hat{\partial}_\sigma\equiv
\Box{}^{-\frac{1}2}\partial_\sigma $ defined in
reference\cite{ak}. Action (\ref{eqb15}) is
\begin{eqnarray}
{S^{(3)L}}_m =\kappa^2 \big<
{k^{Tt}}_{\mu\nu}(\Box{}-m^2){k^{Tt}}^{\mu\nu}-{\theta^{T}}_\mu(\Box{}-2m^2){\theta^{T}}^\mu
-2 \epsilon^{\sigma\alpha\beta}\partial_\alpha {v^{T}}_\beta
\Box{}^{\frac{1}2}{\theta^{T}}_\sigma\nonumber \\ - {v^{T}}_\mu
\Box{}{v^{T}}^\mu
+2v\Box{}v+2\phi\Box{}\phi+4m^2\psi\phi+6m^2\phi^2\big> \,\, .
\label{d1}
\end{eqnarray}
A new transverse variable, ${a^T}_\mu$ is introduced through
\begin{equation}
{\theta^{T}}_\mu \equiv
{\epsilon_\mu}^{\alpha\beta}\hat{\partial}_\alpha
{a^{T}}_\beta\,\, , \label{d2}
\end{equation}
and the action (\ref{d1}) is rewritten as
\begin{eqnarray}
{S^{(3)L}}_m =\kappa^2 \big<
{k^{Tt}}_{\mu\nu}(\Box{}-m^2){k^{Tt}}^{\mu\nu}-{a^{T}}_\mu(\Box{}-2m^2){a^{T}}^\mu
-2 {a^{T}}_\mu\Box{}{v^{T}}^\mu \nonumber \\ - {v^{T}}_\mu
\Box{}{v^{T}}^\mu
+2v\Box{}v+2\phi\Box{}\phi+4m^2\psi\phi+6m^2\phi^2\big> \,\, .
\label{d3}
\end{eqnarray}
The field equations are
\begin{equation}
(\Box{}-m^2){k^{Tt}}_{\mu\nu}=0 \,\, , \label{d4}
\end{equation}
\begin{equation}
\Box{} {v^{T}}_\mu=0 \,\, , \label{d5}
\end{equation}
\begin{equation}
\Box{}v=0\,\, , \label{d6}
\end{equation}
\begin{equation}
{a^{T}}_\mu=0 \,\, , \label{d7}
\end{equation}
\begin{equation}
 \psi=\phi=0 \,\, , \label{d8}
\end{equation}
and reduced action is
\begin{eqnarray}
{S^{(3)L*}}_m =\kappa^2 \big<
{k^{Tt}}_{\mu\nu}(\Box{}-m^2){k^{Tt}}^{\mu\nu}+2v\Box{}v-
{v^{T}}_\mu \Box{}{v^{T}}^\mu \big> \,\, , \label{d9}
\end{eqnarray}
saying that the contortion propagates two massive helicities $\pm
2$, one massless spin-$0$ and two massless ghost vectors. Then,
there is not positive definite Hamiltonian. This observation can
be confirmed in the next section when we will write down the
Hamiltonian density and a wrong sign appears in the kinetic part
corresponding to the canonical momentum of $v_i$ (see eq.
(\ref{eqb33})).

\section{Gauge variance at the linearized regime}

The quadratical Lagrangian density dependent in torsion and
presented in (\ref{eqa3}), has been constructed without free
parameters, with the exception of $m^2$, of course. It has a
particular shape which only gives mass to the spin 2 component of
the contortion, as we see in the perturbative regime. Let us
comment about de non existence of any possible ''residual'' gauge
invariance of the model. The answer is that the model lost its
gauge invariance and it can be shown performing the study of
symmetries through computation of the gauge generator chains. For
this purpose, a $2+1$ decomposition of (\ref{eqb15}) is performed,
this means
\begin{eqnarray}
{S^{(3)L}}_{m} =\kappa^2
\big<[-\dot{k}_{0i}+2\partial_ik_{00}-2\partial_nk_{ni}-2\epsilon_{in}\dot{v}_n
+2\epsilon_{in}\partial_n v_0
]\dot{k}_{0i}+\dot{k}_{ij}\dot{k}_{ij}\nonumber \\
+[2\epsilon_{nj}\partial_nk_{00}+2\epsilon_{nj}\partial_mk_{nm}-\dot{v}_j-2\partial_jv_0]\dot{v}_j+2(\dot{v}_0)^2
+k_{00}\Delta k_{00}\nonumber \\-2k_{0i}\Delta k_{0i}+k_{ij}\Delta
k_{ij}-(\partial_ik_{i0})^2+\partial_nk_{ni}\partial_mk_{mi}
-2\epsilon_{ij}\partial_i v_j
\partial_nk_{n0}\nonumber \\-2\epsilon_{lm}\partial_m v_0
\partial_nk_{nl}+v_0\Delta v_0-v_i\Delta v_i +(\partial_nv_n)^2\nonumber \\
+m^2[2k_{0i}k_{0i}-k_{ij}k_{ij}-2k_{00}k_{ii}+(k_{ii})^2]\big>\,\,,
\label{eqb22}
\end{eqnarray}
where $\Delta\equiv
\partial_i\partial_i$.

Next, the  momenta are
\begin{equation}
\Pi\equiv \frac{\partial\mathcal{L}}{\partial\dot{k}_{00}}=0\,\, ,
\label{eqb23}
\end{equation}
\begin{equation}
\Pi^i\equiv
\frac{\partial\mathcal{L}}{\partial\dot{k}_{0i}}=-2\dot{k}_{0i}-2\epsilon_{in}
\dot{v}_n+2\partial_ik_{i0}-2\partial_nk_{ni}+2\epsilon_{in}
\partial_nv_0\,\, , \label{eqb24}
\end{equation}
\begin{equation}
\Pi^{ij}\equiv
\frac{\partial\mathcal{L}}{\partial\dot{k}_{ij}}=2\dot{k}_{ij}\,\,
, \label{eqb25}
\end{equation}
\begin{equation}
P\equiv
\frac{\partial\mathcal{L}}{\partial\dot{v}_0}=4\dot{v}_0\,\, ,
\label{eqb26}
\end{equation}
\begin{equation}
P^j\equiv
\frac{\partial\mathcal{L}}{\partial\dot{v}_j}=-2\epsilon_{nj}\dot{k}_{0n}-2\dot{v}_j+2\epsilon_{nj}\partial_nk_{00}
+2\epsilon_{nj}\partial_mk_{mn} -2\partial_jv_0\,\, ,
\label{eqb27}
\end{equation}
and we establish the following commutation rules
\begin{equation}
\{k_{00}(x),\Pi (y)\}=\{v_0(x),P(y)\}=\delta^2(x-y)\,\, ,
\label{eqb28}
\end{equation}
\begin{equation}
\{k_{0i}(x),\Pi^j (y)\}=\{v_i(x),P^j(y)\}={\delta^j}_i
\delta^2(x-y)\,\, , \label{eqb29}
\end{equation}
\begin{equation}
\{k_{ij}(x),\Pi^{nm}(y)\}=\frac{1}2({\delta^n}_i{\delta^m}_j+{\delta^m}_i{\delta^n}_j)
\delta^2(x-y)\,\,. \label{eqb30}
\end{equation}

It can be noted that (\ref{eqb23}) is a primary constraint that we
name
\begin{equation}
G^{(K)}\equiv\Pi\,\, , \label{eqb31}
\end{equation}
where $K$ means the initial index corresponding to a possible
gauge generator chain, provided by the algorithm developed in
reference\cite{c}. Moreover, manipulating (\ref{eqb25}) and
(\ref{eqb27}), other primary constraints appear
\begin{equation}
G^{(K)}_i\equiv \partial_n
k_{ni}-\epsilon_{in}\partial_nv_0-\frac{\epsilon_{in}}{4}P^n+\frac{1}{4}\Pi^i\,\,
, \label{eqb32}
\end{equation}
and we observe that $G^{(K)}$ and $G^{(K)}_i$ are first class.

The preservation of constraints requires to obtain the Hamiltonian
of the model. First of all, the Hamiltonian density can be written
as
$\mathcal{H}_0=\Pi^i\dot{h}_{0i}+\Pi^{ij}\dot{h}_{ij}+P\dot{v}_0+P^i\dot{v}_i-\mathcal{L}$,
in other words
\begin{eqnarray}
\mathcal{H}_0 =\frac{\Pi^{ij}\Pi^{ij}}{4}+ \frac{P^2}{8}
-\frac{P^iP^i}{4}+\epsilon_{nj}\partial_mk_{nm}P^j+v_0[\partial_iP^i+4\epsilon_{ml}\partial_m\partial_nk_{nl}]\nonumber
\\+k_{00}[2\partial_m\partial_nk_{mn}-\epsilon_{nm}\partial_nP^m+2m^2k_{ii}]
+2k_{0i}\Delta k_{0i}-k_{ij}\Delta k_{ij}\nonumber
\\+(\partial_ik_{i0})^2-2\partial_nk_{ni}\partial_mk_{mi}
+2\epsilon_{ij}\partial_i v_j
\partial_nk_{n0}+v_i\Delta v_i -(\partial_nv_n)^2\nonumber \\
-m^2[2k_{0i}k_{0i}-k_{ij}k_{ij}+(k_{ii})^2]\,\,. \label{eqb33}
\end{eqnarray}

Then, the Hamiltonian is $H_0=\int dy^2 \mathcal{H}_0(y)\equiv
\big<\mathcal{H}_0\big>_y$ and the preservation of $G^{(K)}$,
defined in (\ref{eqb31}) is
\begin{equation}
\{G^{(K)}(x),H_0\}=-2\partial_m\partial_nk_{mn}(x)+\epsilon_{nm}\partial_nP^m(x)-2m^2k_{ii}(x)\,\,
. \label{eqb34}
\end{equation}
The possible generators chain is given by the rule:
''$G^{(K-1)}+\{G^{(K)}(x),H_0\}=${\it combination of primary
constraints}'', then
\begin{eqnarray}
G^{(K-1)}(x)=2\partial_m\partial_nk_{mn}(x)-\epsilon_{nm}\partial_nP^m(x)+2m^2k_{ii}(x)
\nonumber \\+\big<a(x,y)G^{(K)}(y)+b^i(x,y)G^{(K)}_i(y)\big>_y\,\,
. \label{eqb35}
\end{eqnarray}

The preservation of $G^{(K)}_i$, defined in (\ref{eqb32}), is
\begin{eqnarray}
\{G^{(K)}_i(x),H_0\}=\frac{\partial_n\Pi^{ni}(x)}{2}-\frac{\epsilon_{in}}{4}\partial_nP(x)+\frac{\epsilon_{in}}{2}\Delta
v_n(x) +\frac{\epsilon_{in}}{2}\partial_n\partial_mv_m(x)\nonumber
\\+\frac{\epsilon_{nm}}{2}\partial_i\partial_nv_m(x)-(\Delta
-m^2)k_{0i}(x) \,\, , \label{eqb36}
\end{eqnarray}
then
\begin{eqnarray}
G^{(K-1)}_i(x)=-\frac{\partial_n\Pi^{ni}(x)}{2}+\frac{\epsilon_{in}}{4}\partial_nP(x)-\frac{\epsilon_{in}}{2}\Delta
v_n(x) -\frac{\epsilon_{in}}{2}\partial_n\partial_mv_m(x)\nonumber
\\-\frac{\epsilon_{nm}}{2}\partial_i\partial_nv_m(x)+(\Delta
-m^2)k_{0i}(x)\nonumber
\\
+\big<a^i(x,y)G^{(K)}(y)+{b^i}_j(x,y)G^{(K)}_j(y)\big>_y\,\, .
\label{eqb37}
\end{eqnarray}

The undefined objects $a(x,y)$, $b^i(x,y)$, $a^i(x,y)$ and
${b^i}_j(x,y)$ in  expressions (\ref{eqb35}) and (\ref{eqb37}),
are functions or distributions. If it is possible, they can be
fixed in a way that the preservation of  $G^{(K-1)}(x)$ and
$G^{(K-1)}_i(x)$ would be combinations of primary constraints.
With this, the generator chains could be interrupted and we simply
take $K=1$. Of course, the order $K-1=0$ generators must be first
class, as every one. Next, we can see that all these statements
depend on the massive or non-massive character of the theory.

Taking a chain with $K=1$, the candidates to generators of gauge
transformation are (\ref{eqb31}), (\ref{eqb32}), (\ref{eqb35}) and
(\ref{eqb37}). But, the only non null commutators are
\begin{eqnarray}
\{G^{(1)}_i(x),G^{(0)}_j(y)\}=\frac{m^2}{4}\eta_{ij}\delta^2(x-y)
\,\, , \label{eqb38}
\end{eqnarray}
\begin{eqnarray}
\{G^{(0)}(x),G^{(0)}_i(y)\}=m^2\big(\partial_i\delta^2(x-y)+\frac{b^i(x,y)}{4}\big)
\,\, , \label{eqb39}
\end{eqnarray}
saying that the system of ''generators'' is not first class.
Moreover, the unsuccessful conditions (in the $m^2\neq 0$ case) to
interrupt the chains, are
\begin{eqnarray}
\{G^{(0)}(x),H_0\}=m^2(\Pi^{nn}(x)-2\partial_nk_{0n}(x)) \,\, ,
\label{eqb40}
\end{eqnarray}
\begin{eqnarray}
\{G^{(0)}_i(x),H_0\}=m^2(\partial_nk_{in}(x)+\partial_ik_{00}(x)-\partial_ik_{nn}(x))
\,\, , \label{eqb41}
\end{eqnarray}
where we have fixed
\begin{eqnarray}
a(x,y)=0 \,\, , \label{eqb42}
\end{eqnarray}
\begin{eqnarray}
b^i(x,y)=-2\partial^i\delta^2(x-y) \,\, , \label{eqb43}
\end{eqnarray}
\begin{eqnarray}
a^i(x,y)=0 \,\, , \label{eqb44}
\end{eqnarray}
\begin{eqnarray}
{b^i}_j(x,y)=0 \,\, . \label{eqb45}
\end{eqnarray}

All this indicates that in the case where $m^2\neq 0$ there is not
a first class consistent chain of generators and, then there is no
gauge symmetry.

However, if we revisit the case $m^2=0$, conditions (\ref{eqb40})
and (\ref{eqb41}) are zero and the chains are interrupted. Now,
the generators $G^{(1)}$, $G^{(1)}_i$, $G^{(0)}$ and $G^{(0)}_i$
are first class. Using (\ref{eqb42})-(\ref{eqb43}), the generators
are rewritten again
\begin{equation}
G^{(1)}\equiv\Pi\,\, , \label{eqb46}
\end{equation}
\begin{equation}
G^{(1)}_i\equiv \partial_n
k_{ni}-\epsilon_{in}\partial_nv_0-\frac{\epsilon_{in}}{4}P^n+\frac{1}{4}\Pi^i\,\,
, \label{eqb47}
\end{equation}
\begin{eqnarray}
G^{(0)}=-\frac{\epsilon_{nm}}{2}\partial_nP^m-\frac{\partial_n\Pi^n}{2}
\,\, , \label{eqb48}
\end{eqnarray}
\begin{eqnarray}
G^{(0)}_i=-\frac{\partial_n\Pi^{ni}}{2}+\frac{\epsilon_{in}}{4}\partial_nP-\frac{\epsilon_{in}}{2}\Delta
v_n
-\frac{\epsilon_{in}}{2}\partial_n\partial_mv_m-\frac{\epsilon_{nm}}{2}\partial_i\partial_nv_m+\Delta
k_{0i}\,\, . \label{eqb49}
\end{eqnarray}

Introducing the parameters $\varepsilon (x)$ and $\varepsilon^i
(x)$, a combination of (\ref{eqb46})-(\ref{eqb49}) is taken into
account in the way that the gauge generator is
\begin{eqnarray}
G(\dot{\varepsilon} , \dot{\varepsilon}^i , \varepsilon ,
\varepsilon^i)=\big<
\dot{\varepsilon}(x)G^{(1)}(x)+\dot{\varepsilon}^i(x)G^{(1)}_i(x)+\varepsilon(x)
G^{(0)}(x) +\varepsilon^i(x)G^{(0)}_i(x)\big>\,\, , \label{eqb50}
\end{eqnarray}
and with this, for example the field transformation rules (this
means,  $\delta (...)= \{(...), G \}$) are written as
\begin{eqnarray}
\delta k_{00}= \dot{\varepsilon}\,\, , \label{eqb51}
\end{eqnarray}
\begin{eqnarray}
\delta k_{0i}=
\frac{\dot{\varepsilon}^i}{4}+\frac{\partial_i\varepsilon}{2}\,\,
, \label{eqb52}
\end{eqnarray}
\begin{eqnarray}
\delta k_{ij}=
\frac{1}{4}(\partial_i\varepsilon_j+\partial_j\varepsilon_i)\,\, ,
\label{eqb53}
\end{eqnarray}
\begin{eqnarray}
\delta v_0= \frac{\epsilon_{nm}}{4}\partial_n\varepsilon_m\,\, ,
\label{eqb54}
\end{eqnarray}
\begin{eqnarray}
\delta v_i= \frac{\epsilon_{in}}{4}\dot{\varepsilon}_n
-\frac{\epsilon_{in}}{2}\partial_n\varepsilon\,\, , \label{eqb55}
\end{eqnarray}
and, redefining parameters as follows: $\varepsilon \equiv 2\xi_0$
and $\varepsilon^i= 4 \xi^i$, it is very easy to see that these
rules match with (\ref{eqb9}) and (\ref{eqb10}), as we expected.

\section{YM-extended formulation}

Here we review a possible  quadratical term family which allows to
eliminate non-unitary propagations in the contortion (torsion)
perturbative regime in $2+1$ dimension, at least in a perturbative
regime. The most general shape of a Lagrangian counter terms set
is
\begin{eqnarray}
{S^{(3)}}_0=\kappa^{2} \big<-\frac 14 \,
{(F^{\mu\nu})^\sigma}_\rho {(F_{\mu\nu})^\rho}_\sigma +
a_1{(F_{\mu\nu})^\sigma}_\rho ({F^\mu}_\sigma )^{\nu\rho}+ a_2
{(F_{\mu\nu})^\sigma}_\rho ({F_\sigma}^\rho )^{\mu\nu}\nonumber
\\ + a_3{(F_{\mu\sigma})^\sigma}_\nu
({F^\mu}_\rho )^{\rho\nu}+a_4{(F_{\mu\sigma})^\sigma}_\nu
({F^\nu}_\rho )^{\rho\mu}+a_5((F_{\mu\nu})^{\mu\nu})^2\big> \,\,
,\,\, \label{eqc1}
\end{eqnarray}
where $a_1$, $a_2$, $a_3$, $a_4$ and $a_5$ are real parameters.

A naive try to reach unitarity at a linearized level consists to
perform a direct matching between the perturbative action coming
from (\ref{eqc1}) and the linearized Hilbert-Einstein one, given
by
\begin{eqnarray}
{S_{HE}}^L =-2\kappa^2 \big<
h_{\mu\nu}{G_L}^{\mu\nu}\big>=\kappa^2 \big<
h_{\mu\nu}\Box{}h^{\mu\nu}+ 2\partial_\mu
h^{\mu\sigma}\partial_\nu {h^\nu}_\sigma +2h\partial_\mu
\partial_\nu h^{\mu\nu}-h\Box{}h\big>\,\, ,\nonumber \\\,\, \label{eqc2}
\end{eqnarray}
where $h_{\mu\nu}$ is the metric perturbation and ${G_L}^{\mu\nu}$
is the linearized Einstein's tensor. Then, under perturbations of
the contortion (torsion), one can use again eq. (\ref{eqb7}), this
time in (\ref{eqc1}). Next, making comparison between this result
and (\ref{eqc2}), a linear equation's system for parameters $a_1$,
$a_2$, $a_3$, $a_4$ and $a_5$ arise and only two of them remain
free (i.e., $a_3\equiv \alpha$ and $a_5\equiv \beta$). This means
that for any $\alpha$ and $\beta$ one can get an unitary
(linearized) theory which contains massless spin 2 in 2+1
dimension, in other words, we demand that linearized version of
(\ref{eqc1}) must be proportional to $\big<
k_{\mu\nu}\Box{}k^{\mu\nu}+ 2\partial_\mu
k^{\mu\sigma}\partial_\nu {k^\nu}_\sigma +2k\partial_\mu
\partial_\nu k^{\mu\nu}-k\Box{}k\big>$. These family
of theories labeled by free parameters $\alpha$ and $\beta$, are
\begin{eqnarray}
{S^{(3)}}_{(\alpha,\beta)}=\kappa^{2} \big<-\frac 14 \,
{(F^{\mu\nu})^\sigma}_\rho {(F_{\mu\nu})^\rho}_\sigma
-(1+\alpha){(F_{\mu\nu})^\sigma}_\rho ({F^\mu}_\sigma
)^{\nu\rho}\nonumber
\\ + (\frac58 +\frac{\alpha}2 +\beta) {(F_{\mu\nu})^\sigma}_\rho
({F_\sigma}^\rho )^{\mu\nu} + \alpha {(F_{\mu\sigma})^\sigma}_\nu
({F^\mu}_\rho )^{\rho\nu}\nonumber
\\-(\frac12 +\alpha +4\beta){(F_{\mu\sigma})^\sigma}_\nu
({F^\nu}_\rho )^{\rho\mu}+\beta((F_{\mu\nu})^{\mu\nu})^2\big> \,\,
.\,\, \label{eqc3}
\end{eqnarray}
Another illustrative shape of this action can be obtained using
again (\ref{e40}) and decomposing the contortion in a
symmetric($K_{\mu\sigma}=K_{\sigma\mu}$)-antisymmetric($V_\beta$)
parts as follows
\begin{equation}
{K^\alpha}_{\mu\beta}\equiv{(K_\mu)^{\alpha}}_\beta={\varepsilon^{\sigma\alpha}}_\beta
K_{\mu\sigma}+{\delta^\alpha}_\mu V_\beta
-g_{\mu\beta}V^\alpha\,\, , \label{eqc3a}
\end{equation}
then, action (\ref{eqc3}) is rewritten in the next manner
\begin{eqnarray}
{S^{(3)}}_{(\alpha,\beta)}=\kappa^{2}
\big<K_{\mu\nu}\Box{}K^{\mu\nu}+ 2\nabla_\mu
K^{\mu\sigma}\nabla_\nu {K^\nu}_\sigma +2K\nabla_\mu \nabla_\nu
K^{\mu\nu}-K\Box{}K\nonumber \\
+\,q^{(4)}(K,V)+p^{(3)}(\Gamma,K,V)\big> \,\, ,\,\, \label{eqc3b}
\end{eqnarray}
where $\nabla_\mu$ is the Levi-Civita derivative and $\Box{}\equiv
\nabla_\mu\nabla^\mu$. In (\ref{eqc3b}), $q^{(4)}(K,V)$ means a
polynomial of fourth order in $K_{\mu\nu}$ and $V_\mu$ and first
order in derivatives of these fields, in other words
\begin{eqnarray}
q^{(4)}(K,V)\equiv -\frac12 \, {(f^{\mu\nu})^\sigma}_\rho {[K_\mu
,K_\nu ]^\rho}_\sigma -2(1+\alpha){(f_{\mu\nu})^\sigma}_\rho
[K^\mu ,K_\sigma ]^{\nu\rho}\nonumber
\\ + (\frac54 +\alpha +2\beta) {(f_{\mu\nu})^\sigma}_\rho
[K_\sigma ,K^\rho ]^{\mu\nu} + 2\alpha
{(f_{\mu\sigma})^\sigma}_\nu [K^\mu ,K_\rho ]^{\rho\nu}\nonumber
\\-(1 +2\alpha +8\beta){(f_{\mu\sigma})^\sigma}_\nu
[K^\nu ,K_\rho ]^{\rho\mu}+2\beta(f_{\mu\nu})^{\mu\nu}[K_\alpha
,K_\beta ]^{\alpha\beta}\nonumber
\\-\frac14\, {[K^\mu,K^\nu ]^\sigma}_\rho {[K_\mu ,K_\nu ]^\rho}_\sigma
-(1+\alpha){[K_\mu,K_\nu ]^\sigma}_\rho [K^\mu ,K_\sigma
]^{\nu\rho}\nonumber
\\ + (\frac58 +\frac{\alpha}2 +\beta) {[K_\mu,K_\nu ]^\sigma}_\rho
[K_\sigma ,K^\rho ]^{\mu\nu} + \alpha {[K_\mu,K_\sigma
]^\sigma}_\nu [K^\mu ,K_\rho ]^{\rho\nu}\nonumber
\\-(\frac12 +\alpha +4\beta){[K_\mu,K_\sigma ]^\sigma}_\nu
[K^\nu ,K_\rho ]^{\rho\mu}+\beta([K_\mu,K_\nu ]^{\mu\nu})^2\,\,
,\,\, \label{eqc3c}
\end{eqnarray}
where ${(K_\mu)^{\alpha}}_\beta$ is evaluated on eq. (\ref{eqc3a})
and $(f_{\mu\nu})_{\alpha\beta}\equiv
2{\varepsilon^\sigma}_{\alpha\beta}\nabla_{[\mu}
K_{\sigma\nu]}+2g_{[\nu\alpha}\nabla_{\mu]}V_\beta-
2g_{[\nu\beta}\nabla_{\mu]}V_\alpha$ (symbol $[\mu\nu]$ means
antisymmetrization). The object $p^{(3)}(\Gamma,K,V)$ is a third
order polynomial in fields and first order in derivatives of these
ones and even though its shape is awful, however it identically
vanishes when Christofell's symbols are null. The expression
(\ref{eqc3b}) clearly explains the demanded behavior of the
perturbative regime in a background with a flat metric provided.

There are two possible massive cases. On one hand, the topological
massive model (\ref{eqa2}) can be considered, which is sensitive
under parity. On the other hand, there is a ''Fierz-Pauli'' model
(\ref{eqa3}), whose mass vanishes when one take a null torsion.
Our main pourpose in this section is to study the classical
consistence of field equations (we asume that the torsionless
limit must be consistent with Einstein's theory), and then
focusing the attention at the massless and topological massive
cases.

In the massless theory with cosmological constant, $\lambda$ in
$2+1$ dimension, we introduce a cosmological term as follows
\begin{eqnarray}
{S^{(3)}}_{(\alpha,\beta,
\lambda)}={S^{(3)}}_{(\alpha,\beta)}+\kappa^{2}
\big<q(\alpha,\beta)\lambda^2\big> \,\, ,\,\, \label{eqc4}
\end{eqnarray}
where $q(\alpha,\beta)$ is a (unknown) real function of family's
parameters. Next, in order to consider classical consistence at
the torsionless regime, we take into account some auxiliary fields
(Lagrange multipliers), $b_{\mu\nu}$ and the action with torsion
constraints is given by
\begin{eqnarray}
{S'^{(3)}}_{(\alpha,\beta,
\lambda)}={S^{(3)}}_{(\alpha,\beta)}+\kappa^{2}
\big<q(\alpha,\beta)\lambda^2\big>+\kappa^2\big<b_{\alpha\beta}\,\varepsilon^{\beta\lambda\sigma}
{(A_\lambda)^\alpha}_\sigma \big>\,\, ,\,\, \label{eqc5}
\end{eqnarray}
where arbitrary variations on fields $b_{\mu\nu}$, obviously
provide the condition ${T^\alpha}_{\mu\nu}=0$. Then, the field
equation coming from variations of connection is
\begin{eqnarray}
\nabla_\mu {(\mathcal{F}^{\mu\nu})^\sigma}_\rho+
b_{\rho\mu}\,\varepsilon^{\mu\nu\sigma}=0\,\, ,\,\, \label{eqc5a}
\end{eqnarray}
where $\mathcal{F}_{\mu\nu}$ is defined in terms of the Yang-Mills
curvature, $F_{\mu\nu}$ in the way
\begin{eqnarray}
{(\mathcal{F}^{\mu\nu})^\sigma}_\rho\equiv
{(F^{\mu\nu})^\sigma}_\rho + (\frac54 +\alpha
+2\beta)[({F_\rho}^\sigma )^{\nu\mu} -({F_\rho}^\sigma )^{\mu\nu}]
 \nonumber
\\+2(1+\alpha) [({F^\mu}_\rho )^{\nu\sigma}-({F^\nu}_\rho
)^{\mu\sigma}] +2 \alpha [({F^\nu}_\lambda )^{\lambda\sigma}
{\delta^\mu}_\rho-({F^\mu}_\lambda )^{\lambda\sigma}
{\delta^\nu}_\rho ]\nonumber
\\+(1 +2\alpha +8\beta)[({F^\sigma}_\lambda
)^{\lambda\mu} {\delta^\nu}_\rho-({F^\sigma}_\lambda
)^{\lambda\nu} {\delta^\mu}_\rho ]\nonumber
\\+2\beta(F_{\lambda\kappa})^{\lambda\kappa}(g^{\mu\sigma}{\delta^\nu}_\rho -g^{\nu\sigma}{\delta^\mu}_\rho
)\,\, ,\,\,\nonumber
\\ \label{eqc5b}
\end{eqnarray}
and now, we can match the YM curvature with the
Riemann-Christoffel one (i. e.,
$(F_{\mu\nu})_{\alpha\beta}=R_{\alpha\beta\nu\mu}$), which
satisfies the well known algebraic properties and Bianchi
identities, recalling as follows
\begin{eqnarray}
Symmetry:\,\,\,\,R_{\alpha\beta\nu\mu}=R_{\nu\mu\alpha\beta}\,\,
,\,\, \label{eqc6a}
\end{eqnarray}
\begin{eqnarray}
Antisymmetry:\,\,\,\,R_{\alpha\beta\nu\mu}=-R_{\beta\alpha\nu\mu}=R_{\beta\alpha\mu\nu}=-R_{\alpha\beta\mu\nu}\,\,
,\,\, \label{eqc6b}
\end{eqnarray}
\begin{eqnarray}
Cyclicity:\,\,\,\,R_{\alpha\beta\nu\mu}+R_{\alpha\mu\beta\nu}+R_{\alpha\nu\mu\beta}=0\,\,
,\,\, \label{eqc6c}
\end{eqnarray}
\begin{eqnarray}
Bianchi\,\,identities:\,\,\,\,\nabla_\sigma
R_{\alpha\beta\nu\mu}+\nabla_\mu
R_{\alpha\beta\sigma\nu}+\nabla_\nu R_{\alpha\beta\mu\sigma}=0\,\,
.\,\, \label{eqc6d}
\end{eqnarray}

In $2+1$ dimension, the curvature tensor can be written in terms
of Ricci's tensor ($R_{\mu\sigma}\equiv
{R^\lambda}_{\mu\lambda\sigma}$) and its trace ($R\equiv
{R^\lambda}_\lambda$) in the way
$R_{\lambda\mu\nu\sigma}=g_{\lambda\nu}R_{\mu
\sigma}-g_{\lambda\sigma}R_{\mu\nu}-g_{\mu
\nu}R_{\lambda\sigma}+g_{\mu\sigma}R_{\lambda\nu}-\frac{R}{2}\,
(g_{\lambda\nu}g_{\mu \sigma}-g_{\lambda\sigma}g_{\mu\nu})$. So,
the object defined in (\ref{eqc5b}) takes the shape
\begin{eqnarray}
(\mathcal{F}_{\sigma\nu})_{\lambda\mu}=(\frac32 +4\beta
)R_{\lambda\mu\nu\sigma}+(1+8\beta)(g_{\mu\nu}R_{\lambda
\sigma}-g_{\mu\sigma}R_{\lambda\nu})\nonumber \\+\,2\beta R
(g_{\lambda\nu}g_{\mu \sigma}-g_{\lambda\sigma}g_{\mu\nu})\,\,
,\,\, \label{eqc7}
\end{eqnarray}
which do not depend on parameter $\alpha$. Moreover, if $\beta$ is
fixed as
\begin{eqnarray}
\beta= -\frac18\,\, ,\,\, \label{eqc7a}
\end{eqnarray}
then, relation (\ref{eqc7}) leads to
\begin{eqnarray}
(\mathcal{F}_{\sigma\nu})_{\lambda\mu}\mid_{\beta=
-\frac18}=R_{\lambda\mu\nu\sigma}-\frac{R}4 (g_{\lambda\nu}g_{\mu
\sigma}-g_{\lambda\sigma}g_{\mu\nu})\,\, ,\,\, \label{eqc7b}
\end{eqnarray}
and this one satisfies all symmetry properties of a curvature,
showing in relations (\ref{eqc6a})-(\ref{eqc6c}) with the
exception of the Bianchi identities, (\ref{eqc6d}). It can be
noted that the trace of (\ref{eqc7b}), this means
${(\mathcal{F}_{\sigma\lambda})^\lambda}_\mu$ is the Einstein's
tensor.

Next, some discussion on the critical value (\ref{eqc7a}) will be
performed when the connection's field equation is taking into
account. With the help of symmetry properties, Bianchi's
identities, and relationship between Riemann-Christoffel and Ricci
tensor, the field equation (\ref{eqc5a}) can be rewritten as
follows
\begin{eqnarray}
(\frac12-4\beta)\nabla_\rho
R_{\nu\sigma}-(\frac32+4\beta)\nabla_\sigma
R_{\nu\rho}+(\frac12+2\beta)g_{\nu\rho}\partial_\sigma R +2\beta
g_{\nu\sigma}\partial_\rho R\nonumber
\\+\,
b_{\rho\mu}\,{\varepsilon^\mu}_{\nu\sigma}=0\,\, ,\,\,
\label{eqc8}
\end{eqnarray}
and with some algebraic computation on this last equation, it can
be shown (for all $\beta$) the next symmetry property
\begin{eqnarray}
b_{\nu\mu}= b_{\mu\nu}\,\, ,\,\, \label{eqc8a}
\end{eqnarray}
and
\begin{eqnarray}
(\beta-\frac58)b_{\mu\nu}= 0\,\, ,\,\, \label{eqc8b}
\end{eqnarray}
\begin{eqnarray}
(\beta+\frac18)\partial_\mu R=0\,\, .\,\, \label{eqc8c}
\end{eqnarray}
Consistence condition (\ref{eqc8b}) stablishes that the work out
of Lagrange multipliers depends on the following restriction
\begin{eqnarray}
\beta\neq \frac58\,\, ,\,\, \label{eqc9}
\end{eqnarray}
then, $b_{\mu\nu}=0$. This last result means that one can
consistently replace $b_{\mu\nu}=0$ inside the action (\ref{eqc5})
and, then the torsionless limit can be recovered through the
condition ${T^\lambda}_{\mu\nu}=0$ imposed on the new field
equations.

Condition (\ref{eqc9}) induces a wide set of possible vacuum's
solutions, including non-Einstenian ones beside (A)dS, because eq.
(\ref{eqc8c}) becomes an identity when it is evaluated on the
critical $\beta$ given by (\ref{eqc7a}). This fact is confirmed
when  $\beta=-\frac18$ is introduced in eq. (\ref{eqc8}), in other
words
\begin{eqnarray}
\nabla_\rho \mathcal{R}_{\nu\sigma}-\nabla_\sigma
\mathcal{R}_{\nu\rho}=0\,\, ,\,\, \label{eqc9a}
\end{eqnarray}
where notation means
\begin{eqnarray}
\mathcal{R}_{\mu\nu}\equiv R_{\mu\nu}- \frac{g_{\mu\nu}}4\,R\,\,
.\,\, \label{eqc9b}
\end{eqnarray}
It can be observed that equation (\ref{eqc9a}) looks like eq.
(\ref{g1a}), but here, as one can expect the trace
$\sigma-\lambda$ of (\ref{eqc9a}) is an identity.

In order to conclude the comments on the massless theory, next we
consider the field equation which comes from variations on metric
of the action (\ref{eqc5}) and it can be written in terms of
Ricci's tensor and Ricci's scalar as follows
\begin{eqnarray}
(\frac32-\alpha+12\beta)R_{\sigma\mu}{R^\sigma}_\nu
-(\frac12-\alpha+6\beta)RR_{\mu\nu}-(1-\alpha+4\beta)R^{\sigma\rho}R_{\sigma\rho}
g_{\mu\nu}\nonumber \\
+\,(\frac{5}{16}-\frac{\alpha}2+2\beta)R^2g_{\mu\nu}+\frac{q}2\,\lambda^2g_{\mu\nu}=0\,\,
.\,\, \label{eqc10}
\end{eqnarray}
Immediately, the consistence with (A)dS solutions is evaluated by
replacing the contractions of $R_{\rho\mu\nu\sigma}= \lambda
(g_{\rho\sigma}g_{\mu\nu}-g_{\rho\nu}g_{\mu\sigma})$  in
(\ref{eqc10}). This gives
\begin{eqnarray}
q(\alpha)=\frac32-4\alpha\,\, ,\,\, \label{eqc10a}
\end{eqnarray}
and this indicates that if $\alpha=\frac38$ is introduced in
action (\ref{eqc5}) get implicit (A)dS solutions from its field
equations.

Now we take a look on the gauge formulation of topological masive
gravity with cosmological constant, considering the YM-extended
action at the torsionless limit, this means
\begin{eqnarray}
S'={S^{(3)}}_{(\alpha,\beta)}+\frac{m\kappa^2}{2}\big<\varepsilon^{\mu\nu\lambda}\,tr\big(
A_\mu\partial_\nu A_\lambda +\frac{2}{3}\, A_\mu A_\nu A_\lambda
\big)\big>+\kappa^{2} \big<q(\alpha)\lambda^2\big>\nonumber
\\ +\kappa^2\big<b_{\alpha\beta}\,\varepsilon^{\beta\lambda\sigma}
{(A_\lambda)^\alpha}_\sigma \big>\,\, ,\,\, \label{eqc11}
\end{eqnarray}
where $q(\alpha)$ is defined by (\ref{eqc10a}) then, this action
is consistent with (A)dS solutions when $m=0$. Variations on the
metric conduce to the known equations (\ref{eqc10}). So, the
connection field equation is
\begin{eqnarray}
\nabla_\mu {(\mathcal{F}^{\mu\nu})^\sigma}_\rho+
\frac{m}2\,\varepsilon^{\alpha\beta\nu}{(F_{\alpha\beta})^\sigma}_\rho
+b_{\rho\mu}\,\varepsilon^{\mu\nu\sigma}=0\,\, ,\,\, \label{eqc12}
\end{eqnarray}
and ${(\mathcal{F}^{\mu\nu})^\sigma}_\rho$ is defined in
(\ref{eqc5b}). Recalling that
$(F_{\mu\nu})_{\alpha\beta}=R_{\alpha\beta\nu\mu}$ in a
torsionless space-time, equation (\ref{eqc12}) can be rewritten in
terms of Ricci's tensor as follows
\begin{eqnarray}
(\frac12-4\beta)\nabla_\rho
R_{\nu\sigma}-(\frac32+4\beta)\nabla_\sigma
R_{\nu\rho}+(\frac12+2\beta)g_{\nu\rho}\partial_\sigma R +2\beta
g_{\nu\sigma}\partial_\rho R\nonumber
\\-\,m{\varepsilon^{\alpha\beta}}_\nu
(g_{\alpha\sigma}R_{\beta\rho}-g_{\alpha\rho}R_{\beta\sigma}-\frac{R}{2}\,
g_{\alpha\sigma}g_{\beta\rho})+
b_{\rho\mu}\,{\varepsilon^\mu}_{\nu\sigma}=0\,\, .\,\,
\label{eqc13}
\end{eqnarray}
Performing some algebraic manipulation on this last equation,
conditions (\ref{eqc8a}) and (\ref{eqc8c}), which establish the
symmetry property of Lagrange multipliers and the indetermination
of scalar curvature when $\beta=-\frac18$, rise again in a similar
way that they do in the massless theory.

Then, using condition (\ref{eqc9}), the Lagrange multipliers are
given by
\begin{eqnarray}
b_{\mu\nu}=2\bigg(\frac{\beta+\frac18}{\beta-\frac58}\bigg)m
R_{\mu\nu}-\bigg(\frac{\beta+\frac38}{\beta-\frac58}\bigg)\frac{mR}2
g_{\mu\nu}\,\, ,\,\, \label{eqc14}
\end{eqnarray}
and if (\ref{eqc7a}) is fixed, the  result (\ref{eqa14a}) is
recovered. So, evaluating the theory on $\beta=-\frac18$, the
action  (\ref{eqc13}) becomes in a similar form as in
(\ref{eq13}), this means
\begin{eqnarray}
\nabla_\mu \mathcal{R}_{\sigma\lambda}-\nabla_\lambda
\mathcal{R}_{\sigma\mu}
-m\,{\varepsilon^{\nu\rho}}_\sigma(g_{\lambda\nu}\mathcal{R}_{\mu\rho}-g_{\mu\nu}\mathcal{R}_{\lambda\rho}
-\frac23\,\mathcal{R}\,g_{\lambda\nu}g_{\mu\rho}) =0\,\, ,\,\,
\label{eqc15}
\end{eqnarray}
where $\mathcal{R}_{\mu\nu}$ is defined as in (\ref{eqc9b}) and
the trace $\sigma-\lambda$ is an identity, as one can expect.

\section{Conclusion}

A perturbative regime based on arbitrary variations of the
contortion and metric as a (classical) fixed background, is
performed in the context of a pure Yang-Mills formulation of the
$SO(1,2)$ gauge group. There, we analyze in detail the physical
content and the well known fact that a variational principle based
on the propagation of torsion (contortion), as dynamical and
possible candidate for a quantum canonical description of gravity
in a pure YM formulation gets serious difficulties.

In the $2+1$ dimensional massless case we show that the theory
contains three massless degrees of freedom, one of them a
non-unitary mode, considering a non dynamical background's metric.
Then, introducing appropiate quadratical terms dependent on
torsion, which preserve parity and general covariance, we can see
that the linearized limit do not reproduces an equivalent pure
Hilbert-Einstein-Fierz-Pauli massive theory for a spin-2 mode and,
moreover there is other non-unitary modes. Roughly speaking, at
first sight one can blame it on the kinetic part of YM formulation
because the existence of non-positive Hamiltonian connected with
non-unitarity problem. Nevertheless there are other possible $F^2$
models (or simply {\it YM-extended}) which could solve the
unitarity problem.

Exploring the massless and the topological massive gravity models
in $2+1$ dimension, the well known existence of a YM-extended
theories family is noted. This family is labeled with two free
parameters, $\alpha$ and $\beta$ and can cure non-unitary
propagations in the linearized level. Moreover, when the classical
consistence between these type of theories and the Einstein's one
is tackled, what we have mentioned as {\it torsionless limit}, the
relationship between parameter $\alpha$ and the shape of the
coupling of the cosmological constant in the action, is shown.

Meanwhile, the parameter $\beta$ get two types of critical values.
On one side, the number $\beta=\frac58$ is connected to the
classical consistence requirement which demands the introduction
of torsion's Lagrangian constraints with solvable Lagrange
multipliers. On the other side, the value $\beta=-\frac18$
establishes a wide set of theories, including the Einstein's
solutions after the imposition of a auxiliary condition
$R=constant$ and non-Einsteinian ones when the Ricci scalar became
an arbitrary function in an empty space-time. But, even though the
Lagrangian extension of the YM formulation for gravity conduces to
the well known fact that there exists unphysical classical
solutions, the same occurs (in a much less severe way) without
these corrections and one can recall the YM pure formulation gives
rise a set of solutions for the massless and topological massive
gravity with the property $R=constant$ and only Einsteinian
results can be obtained if the auxiliary condition $R=-6\lambda$
is fixed.

A generalization of the study of physical content of the
YM-extended model in a non perturbative level, including a
dynamical metric would be considered elsewhere.

\section*{Acknowledgments}

Author thanks referee's remarks. This work is partially supported
by FACYT-UC.

\end{document}